\documentclass[sigconf,nonacm]{acmart}
\usepackage[frozencache]{minted}

\usepackage{enumitem}

\usepackage{csvsimple}

\usepackage[]{algorithm2e}
\usepackage{mathrsfs}

\usepackage{tcolorbox}
\usepackage{mathtools}

\usepackage[disable]{todonotes}

\usepackage{amsthm}

\usepackage{commath}
\usepackage{comment}
\usepackage{multirow}
\usepackage{pgfplots, pgfplotstable, booktabs, colortbl, siunitx, array}
\usepackage{longtable}
\usepackage{subcaption}
\usepackage{datatool}
\usepackage{xspace}
\usepackage{bm}  
\newcommand{\diff}{\mathrm{d}}
\usepackage{lineno}  

\newcolumntype{R}[1]{>{\raggedleft\arraybackslash }b{#1}}

\newcommand{\mfT}{\mathfrak{T}}

\newcommand{\martin}[1]{\medskip \todo[inline,color=purple!40,caption={}]{Martin: #1} \medskip}

\newcommand{\mbtgym}{\texttt{mbt\_gym\xspace}}

\newcommand\srccode[3]{
\inputminted[frame=single,autogobble,fontsize=\scriptsize,firstline=#2,lastline=#3]{python}{#1}
}

\AtBeginDocument{%
  \providecommand\BibTeX{{%
    \normalfont B\kern-0.5em{\scshape i\kern-0.25em b}\kern-0.8em\TeX}}}

\acmConference[ICAIF'22]{International Conference on AI in Finance}{Nov 02--04,
  2022}{New York}
\acmPrice{15.00}
\acmISBN{978-1-4503-XXXX-X/18/06}

\begin{document}

\title{Model-based gym environments for limit order book trading}


\author{Joseph Jerome}
\affiliation{%
  \institution{Dept.\ of Computer Science, University of Liverpool}
   \city{Liverpool}
   \country{UK}
}
\authornote{Authors contributed equally to the paper.}
\email{j.jerome@liverpool.ac.uk}

\author{Leandro S\'anchez-Betancourt}
\affiliation{%
  \institution{Dept.\ of Mathematics, King's College London}
   \city{London}
   \country{UK}
}
\authornotemark[1]
\email{leandro.sanchez-betancourt@kcl.ac.uk}

\author{Rahul Savani}
\affiliation{%
	\institution{Dept.\ of Computer Science, University of Liverpool}
	\city{Liverpool}
	\country{UK}
}
\authornotemark[1]
\email{rahul.savani@liverpool.ac.uk}

\author{Martin Herdegen}
\affiliation{%
  \institution{Dept.\ of Statistics, University of Warwick}
   \city{Coventry}
   \country{UK}
}
\email{m.herdegen@warwick.ac.uk}


\begin{abstract}

Within the mathematical finance literature there is a rich catalogue of mathematical models
for studying algorithmic trading problems -- such as market-making and optimal execution -- in
limit order books.
This paper introduces \mbtgym, a Python module that provides a
suite of gym environments for training reinforcement learning (RL) agents to
solve such model-based trading problems. 
The module is set up in an extensible way to allow the combination of 
different aspects of different models.
It supports highly efficient implementations of vectorized environments to allow
faster training of RL agents.
In this paper, we motivate the challenge of using RL to solve
such model-based limit order book problems in mathematical finance, we explain
the design of our gym environment, and then demonstrate its use in solving 
standard and non-standard problems from the literature.
Finally, we lay out a roadmap for further development of our module, which we
provide as an open source repository on GitHub so that it can serve as a focal
point for RL research in model-based algorithmic trading.
\end{abstract}



\keywords{limit order book, market-making, optimal execution, liquidity provision, inventory risk, reinforcement learning}

\maketitle

\section{Introduction}

A substantial proportion of financial markets use the limit order book (LOB)
mechanism to match buyers and sellers~\cite{gould2013limit}.
Consequently, LOBs have been a major object of study within mathematical 
finance.
A wide range of mathematical models that capture price dynamics and order
arrivals have been developed, and then trading problems, such as market making
or optimal execution, have been analysed for these models.
The models are differentiated primarily by the stochastic processes that drive
the price and order dynamics, the actions available to the agent, and the agent's reward function.

Typically these problems have been solved (often approximately) using standard
methods from the theory of Partial Differential Equations (PDEs), namely by formulating
the Hamilton-Jacobi-Bellman (HJB) equation and using Euler schemes or finite-difference methods to solve them numerically. 
However:

\begin{enumerate}[leftmargin=0.5cm]
\itemsep0mm

\item The numerical HJB approach requires solution schemes that are
	tailored to the particular stochastic processes of the model.
	Ideally, solution approaches would be more \emph{model agnostic}.

\item The numerical HJB approach suffers particularly strongly from the curse of dimensionality,
	which has also constrained the types of models that have been considered
	and solved this way.

\item When employing the HJB approach, one often looks for semi-explicit solutions. For this reason, one considers `nice' functions to define models, which constrains the range of trading problems that are considered in the literature.

\martin{I am still not fully happy with (3) - the control being in feedback form always follows in the HJB approach. I would rather write something like this:
When employing the HJB approach, one often looks for semi-explicit solutions. For this reason, one considers 'nice' functions as model input, which constrains the range of trading problems that are considered in the literature.}

\end{enumerate}

Reinforcement Learning (RL) is an approach to solving control problems that 
is based on trial and error.
It has seen very rapid development since the Deep Learning revolution, with
a number of prominent success stories.
We contend that using RL to solve model-based LOB trading problems is important
and exciting:
\begin{itemize}[leftmargin=0.3cm]
\item for mathematical finance by providing a complementary solution method in
	addition to PDE approaches that will enable the solution of richer and more
	realistic models; and
\item for the RL community by providing a new class of problems on which
	to develop and understand 
        different RL algorithms.
\end{itemize}

In relation to the highlighted weaknesses of the HJB approach, we note that by
using model-free RL, one can in principle solve richer models with
assumptions that do not suit the HJB approach well, use the same learning method
across many models, and solve higher-dimensional and thereby richer models.
We present an open-source benchmark module that implements a
range of LOB models and trading problems from the mathematical finance
literature, along with RL-based solution methods. Our goal is to showcase
the potential of RL for solving these types of problems, and facilitate further
research in this exciting direction.

While RL is very promising for solving these problems, RL does have a major
accepted weakness of being very sample inefficient. Fortunately, the model-based
trading problems that we study allow for highly efficient implementations of
vectorized environments, which we leverage in our implementations, and we find
that using this vectorization (essentially parallelization) is crucial to get
close-to-optimal solutions to the benchmark problems that we study.

\smallskip

\noindent
\textit{Our main contributions.}
\begin{itemize}[leftmargin=0.3cm]

\item We provide an open source repository of unified gym environments for
	a range of model-based LOB trading problems.

\item We provide optimal baseline agents so that one can benchmark
	the performance of RL algorithms.

\item A key strength of our implementations is the modular design that supports
	picking and choosing different components of existing models, for example
	using a reward function from one model, with the prices and execution
	processes from another.

\item We have connected our environments to Stable	Baselines 3 (SB3), which
provides a suite of state-of-the-art RL algorithms. These algorithms perform robustly across a variety of tasks, allowing a ``plug-and-play'' approach to training RL agents on previously unstudied problems. We demonstrate its use by quickly solving a popular market making problem to near optimality using proximal policy optimisation \cite{SchulmanWDRK17}, showing the benefit of our custom highly-efficient
	vectorized environments.
	
\item Going further, we lay out a roadmap for further developments of
	our RL for model-based trading benchmark suite, both in terms of the models
	and trading problems and in terms of the RL algorithms used to find solutions.

\end{itemize}

\noindent
The module is available at: \url{https://github.com/JJJerome/mbt_gym}.







\subsection{Literature review}

\subsubsection{Mathematical finance}
The market-making problem with limit orders was first introduced in \cite{ho1981optimal} and  mathematically formalized three decades later by \cite{avellaneda2008high}. From that point on, a large volume of research has been produced by the mathematical finance community. The work of \cite{gueant2013dealing} presents an explicit solution to the market making problem with inventory constraints. Examples of subsequent developments featuring more realistic models are those  by \cite{Guilbaud2011,CJR14}. Other relevant extensions include the works of \cite{CDJ2018,cartea2020market}
where signals are incorporated into the framework for both 
continuous action spaces and 
binary actions.  The community continues  to work on aspects of the market making problem, from options \cite{baldacci2021algorithmic} and foreign exchange \cite{barzykin2022dealing} to automated market making~\cite{cartea2022decentralised}. 

Similar to the market making case, the optimal execution literature is vast; starting with \cite{bertsimas1998optimal,almgren2001optimal}, the problem has attracted increasing attention. A significant focus of model design is on how the market processes the order flow from the liquidator (and other market participants), and how signals can be constructed and exploited \cite{cartea2016incorporating,cartea2015optimal,gatheral2012transient,forde2022optimal,neuman2022optimal,donnelly2020optimal,kalsi2020optimal,cartea2020optimal}. 
Further investigations include the  case of stochastic volatility and liquidity \cite{almgren2012optimal}, 
stochastic price impact \cite{barger2019optimal,fouque2022optimal}, and latency and related frictions \cite{moallemi2013or,cartea2021optimal}.

See~\cite{cartea2015algorithmic,gueant2016financial} for an excellent textbook treatment of both 
optimal execution and market making.


\subsubsection{Reinforcement learning for high-frequency trading literature}
The following section provides a brief summary of the reinforcement learning for high-frequency trading 
literature, paying particular attention to the dynamics of the learning environment since that is the main focus of this paper. This section also mainly focuses on the market-making problem as opposed to the optimal execution problem since the majority of the environments currently provided in \mbtgym~focus on market-making, however, a brief summary of the RL for optimal execution literature is also included.

There are broadly three main approaches to modelling the market dynamics for a reinforcement learning environment. The first is to use a ``market replay'' approach, in which historical data is replayed in a simulator, and the learning agent interacts with it. The second is model-based approaches, a category in which this paper sits. The final approach is that of agent-based market simulators in which hand-crafted agents that aim to model real-life market participants are allowed to interact with each other and the training agent through an exchange mechanism. A good summary of the merits of each approach is provided in \citet{Balch19}.

Papers on high-frequency trading which train agents using a market replay simulator include: \citet{Nevmyvaka2006} and \citet{jerome2022market}, which use Nasdaq data; \citet{spooner2018market} which uses data from LSE; \citet{ZBW20} which uses futures data from CME; \citet{XCH22performance}, which uses data from XSHE; and \citet{patel2018optimizing}, \citet{Sadighian19} and \citet{GasperovK21}, which uses cryptocurrency data.

Some notable papers that consider the market-making problem in a model-based environment include: \citet{chan2001electronic}, which uses a Poisson-process-driven model similar to \citet{glosten1985bid}; \citet{Kim02}, who fit a hidden Markov model to Nasdaq trade data; \citet{LG18}, which considers the market making problem in the Poisson-process-driven model of \citet{ContST10}; and \citet{spooner2020robust}, which considers a robust adversarial version of the problem in \cite[Section 10.2]{cartea2015algorithmic}.

Finally, agent-based high-frequency trading simulators are used to train RL agents in \citet{ganesh2019reinforcement}, \citet{Kum20deep}, \citet{KFMW20} and \citet{AMV+21}.

\section{Design of the module}

Our module has been designed with the following principles in mind.
The code should be extensible and general and minimize code duplication.
For example, all our gym environments inherit from a class
\texttt{TradingEnvironment} and use the same following shared \texttt{step}
and ``update'' functions.
The \texttt{step} function, shown below, is not specific to the trading 
problem.
\srccode{src_code/TradingEnvironment.py}{88}{96}
\noindent
\texttt{\_update\_state}, shown next, is specific to the trading
problem.
\srccode{src_code/TradingEnvironment.py}{104}{113}

We are able to use a single \texttt{\_update\_state} function across many 
standard LOB trading models from the literature because they all share the 
same high-level dependency order between the different stochastic processes.
In particular, this means that we can use a fixed sequence for computing
the next step in these stochastic processes, with the order arrivals process updating first and then driving the mid-price process and then the fill probability model.
The function \texttt{\_update\_market\_state}, shown next, deals with simulating these.
\pagebreak

\srccode{src_code/TradingEnvironment.py}{115}{127}

\noindent
Finally, \texttt{\_update\_agent\_state} deals with updating
the cash and inventory of the agents due to executions and price movements,
and depends on the action space of the problem.

\srccode{src_code/TradingEnvironment.py}{129}{150}

\subsection{Arrivals processes}

\paragraph{Poisson process}
The class \texttt{PoissonArrivalModel}	presents the  most frequently used model for arrival of order flow in the market-making literature. Here, the arrival of orders 
from market participants to buy/sell is modelled with a Poisson process $(M^\pm_t)_{t\geq 0}$ with intensities $\lambda^\pm\in\mathbb{R}^+$.

\paragraph{Hawkes process}
The  self-exciting mechanism of the Hawkes processes is coded in the \texttt{HawkesArrivalModel} class -- see Appendix A.3 in \cite{cartea2015algorithmic}. Here,  for simplicity, we  focus on Hawkes processes with exponential kernels. Let $(\lambda^\pm_t)_{t\geq 0}$ be the stochastic intensities for buy/sell order flow and let $(M^\pm_t)_{t\geq 0}$ be the associated counting processes. Then, we implement the slightly more general stochastic intensities 
\begin{equation}
    \diff \lambda^\pm_t = \kappa\,\left(\bar{\lambda} - \lambda^\pm_t\right)\,\diff t +\gamma \, \diff M^\pm_t \,,
\end{equation}
where $\kappa > 0$ is the mean reversion speed,\footnote{Note that $\kappa^\alpha$ needs to be sufficiently large for the Hawkes process to be stationary.} $\bar{\lambda}>0$ is the baseline arrival rate, and $\gamma$ is the jump size parameter.

\subsection{Mid-price processes}
In what follows $(W_t)_{t\geq 0}$ is a standard Brownian motion. We separate the implemented mid-price models into two groups: (i) \textit{Brownian motion} mid-price models and (ii)  \textit{mean-reverting drift dynamics}.

\paragraph{Brownian motion}  The \texttt{BrownianMotionMidpriceModel} class describes the well-known arithmetic Brownian motion; here, the mid-price can be written as 
\begin{equation}
    S_t = S_0 + \mu\,t+ \sigma\,W_t\,,
\end{equation}
where $\mu\in \mathbb{R}$ is the drift parameter and $\sigma\in\mathbb{R}^+$ is the volatility. Another implemented model under this category is the\\ \texttt{GeometricBrownianMotionMidpriceModel} that is the well-known strong solution to the stochastic differential equation (SDE) 
\begin{equation}
    \diff S_t = \mu\,S_t\,\diff t+ \sigma\,S_t\,\diff W_t\,.
\end{equation}

Lastly, we implemented \texttt{BrownianMotionJumpMidpriceModel} which features in \cite[Section 5.2]{gueant2013dealing}, where the mid-price incorporates the impact of the filled orders, i.e., 
\begin{equation}
    S_t = S_0 + \sigma\, W_t - \xi^-\, M^-_t + \xi^+\,M^+_t\,,
\end{equation}
where $M^\pm_t$ are the buy/sell market orders from liquidity takers and $\xi^\pm \in\mathbb{R}^+$ are the permanent price impact parameters.

\paragraph{Mean-reverting drift dynamics}
The \texttt{OuMidpriceModel} class accommodates models where the mid-price process follows  Ornstein--Uhlenbeck (OU) dynamics -- examples in the algorithmic trading literature include \cite{bergault2022multi,cartea2021deep}.

This class enables us to define \texttt{OuDriftMidpriceModel} that implements popular models where the mid-price process has a short-term alpha signal given by an OU process -- see \cite{lehalle2019incorporating,micheli2021closed,cartea2016incorporating}.

More precisely, in such models the mid-price process follows 
\begin{equation}\label{eq: mid price alpha}
    \diff S_t = \alpha_t\,\diff t + \sigma^S\,\diff W^S_t\,,
\end{equation}
where $\sigma^S>0$ is the volatility of the mid-price process, $W^S$ is a Brownian motion, and  $(\alpha_t)_{t\geq 0}$ is an OU process satisfying 
\begin{equation}
\diff\alpha_t = \kappa^\alpha (\bar{\alpha} - \alpha_t)\,\diff t +\sigma^\alpha\,\diff W^\alpha_t.
\end{equation}
Here, $\sigma^\alpha>0$ is the volatility of the OU process, $\kappa^\alpha\geq 0$ is the mean-reversion speed, $\bar{\alpha}$ is the mean-reversion level, and $W^\alpha$ is an independent Brownian motion.

Similar to the example above, \texttt{OuJumpMidpriceModel} encompasses dynamics such as 
\begin{equation}
    \diff\alpha_t = \kappa^\alpha (\bar{\alpha} - \alpha_t)\,\diff t +\sigma^\alpha\,\diff W^\alpha_t  - \xi^-\, M^-_t + \xi^+\,M^+_t\,,
\end{equation}
and helps define the class  \texttt{OuJumpDriftMidpriceModel} which is the mid-price \eqref{eq: mid price alpha} that features in \cite{CJR14}.

\subsection{Fill probability models}

Most models in the literature use \texttt{ExponentialFillFunction} as the fill probability model. Here, the probability that an order posted at depth $\delta^\pm$ is filled is given by
$\mathbb{P}[\textrm{fill}|\delta^\pm] = e^{-\kappa\,\delta^\pm}$,
for a fill exponent $\kappa>0$ -- observe that if $\delta^\pm\in\mathbb{R}^+$, the resulting value lies in $[0,1]$.

Given that the fill probability model is an instance of the class \texttt{StochasticProcess}, this permits an exponential fill probability model with a stochastic $\kappa$ that is affected by arrivals, actions, and fills. Other functional forms are straightforward generalizations. In particular, we implemented \texttt{TriangularFillFunction}  which receives as in input a parameter $\delta_{\texttt{max}}>0$ 
(\texttt{max\_fill\_depth} in the code base), and defines the probability of a fill as one if $\delta^\pm<0$, zero if $\delta^\pm>\delta_{\texttt{max}}$, and $1-\delta^\pm/\delta_{\texttt{max}}$ otherwise. This is a natural way of defining the fill probability function, but we are unaware of its use in the literature.

Similar to the previous two cases, our \texttt{PowerFillFunction} class -- which features in \cite{CJR14}, takes as an input a \texttt{fill\_exponent} $\alpha>0$ and a \texttt{fill\_multiplier} $\kappa>0$, and, for $\delta^\pm\geq 0$ defines the probability of a fill as 
\begin{equation}
    \frac{1}{1+ \left(\kappa\,\delta^\pm\right)^\alpha}\,.
\end{equation}

All fill probability models have a \texttt{max\_depth} property that helps to constrain the range of the action space. In particular, this value is chosen to be the value of $\delta^\pm$ such that the probability of not getting filled when posting at $\delta^\pm$ is less than 1\%.

\subsection{Action spaces}
We implemented three ways in which the agent interacts with the LOB, and we introduce them next.

Firstly, the type \texttt{limit} is one where the market partipant decides to be posted at distance $\delta^+$ from the mid-price on the ask side, and at distance $\delta^-$ from the mid-price on the bid-side. Most of the market-making literature and a significant portion of the optimal execution literature fall within this category.

Secondly, type \texttt{limit-and-market} contains the previous case (i.e., \texttt{limit}) and allows the agent to also send market orders for a single unit of the asset. The market orders can be either to buy or to sell (or both) and it is assumed that they would be executed at distance \texttt{minimum\_tick\_size} from the mid-price in the respective side of the book that the market order was targeting.

Lastly, the action type \texttt{touch} encodes models where the market maker decides whether or not to be posted at the best quotes. The action space is binary and requires
\texttt{minimum\_tick\_size} which records the distance at which the best quotes lie from the mid-price. Upon the arrival of a liquidity taking order to buy/sell, the agent's limit order gets filled if they were posted at the best ask/bid.

\subsection{Reward functions}
\martin{The reward function PnL is not used anymore. (Maybe we can omit it.) Also, for the Running Inventory Aversion I would not subtract Y_0 - or if you want to keep this also subract Y_0 in the exponential term. Otherwise this is inconsistent.}

Let $T>0$ and $\mfT = [0,T]$; in this section we use $(X_t)_{t\in\mfT}$ for the cash process of the trader, $(Q_t)_{t\in\mfT}$ for the inventory, and $(S_t)_{t\in\mfT}$ the midprice process. The reward function \texttt{PnL} (for ``profit and loss'', aka risk-neutral reward) captures the changes in the mark-to-market value $Y_t := X_t + Q_tS_t$ of the trader's position. Specifically, the \texttt{PnL} can be written as $Y_T - Y_0$.

The reward \texttt{RunningInventoryPenalty} receives two parameters: the \texttt{per\_step\_inventory\_aversion}, denoted by  $\phi\geq 0$, and the \texttt{terminal\_inventory\_aversion}, denoted by $a\geq 0$. The reward is then defined as the time-discretized version of
\begin{equation}
Y_T - Y_0 - a\,Q_T^2 -\phi\int_0^T Q_t^2 dt\,.
\end{equation}
Lastly, the \texttt{ExponentialUtility} reward takes a \texttt{risk\_aversion} parameter $\gamma>0$ and computes $\exp\left(-\gamma\left(Y_T-Y_0\right)\right)$.
 
\subsection{Examples of currently supported models}

To give a sense of the range of models that we have implemented as extensions 
of this base trading class, we give examples, first, in
Table~\ref{tab:standard}, of standard models 
from the literature, and then, in Table~\ref{tab:hybrid} of new hybrid models that become immediately available
to use by combining components from different standard models.

\subsection{Vectorized environments}
A key feature of the gym environments in \mbtgym~is their highly parallelized nature. Single trajectories of states visited and rewards received for a given policy necessarily have very high variance. This is due to the stochasticity coming from many different places: first, there is the stochasticity of the midprice process; second, there is randomness in the arrival process; third, there is randomness in whether the agent's limit orders are filled or not.

The standard approach to parallelize reinforcement learning ``rollouts'' is to spin up many environments across multiple threads or CPUs and let each generate trajectories. On a single machine, this multiprocessing can be done using the concurrent.futures package. These trajectories are then sent to a central location for training.

\begin{figure}[h]
\centering
\includegraphics[width=0.7\linewidth]{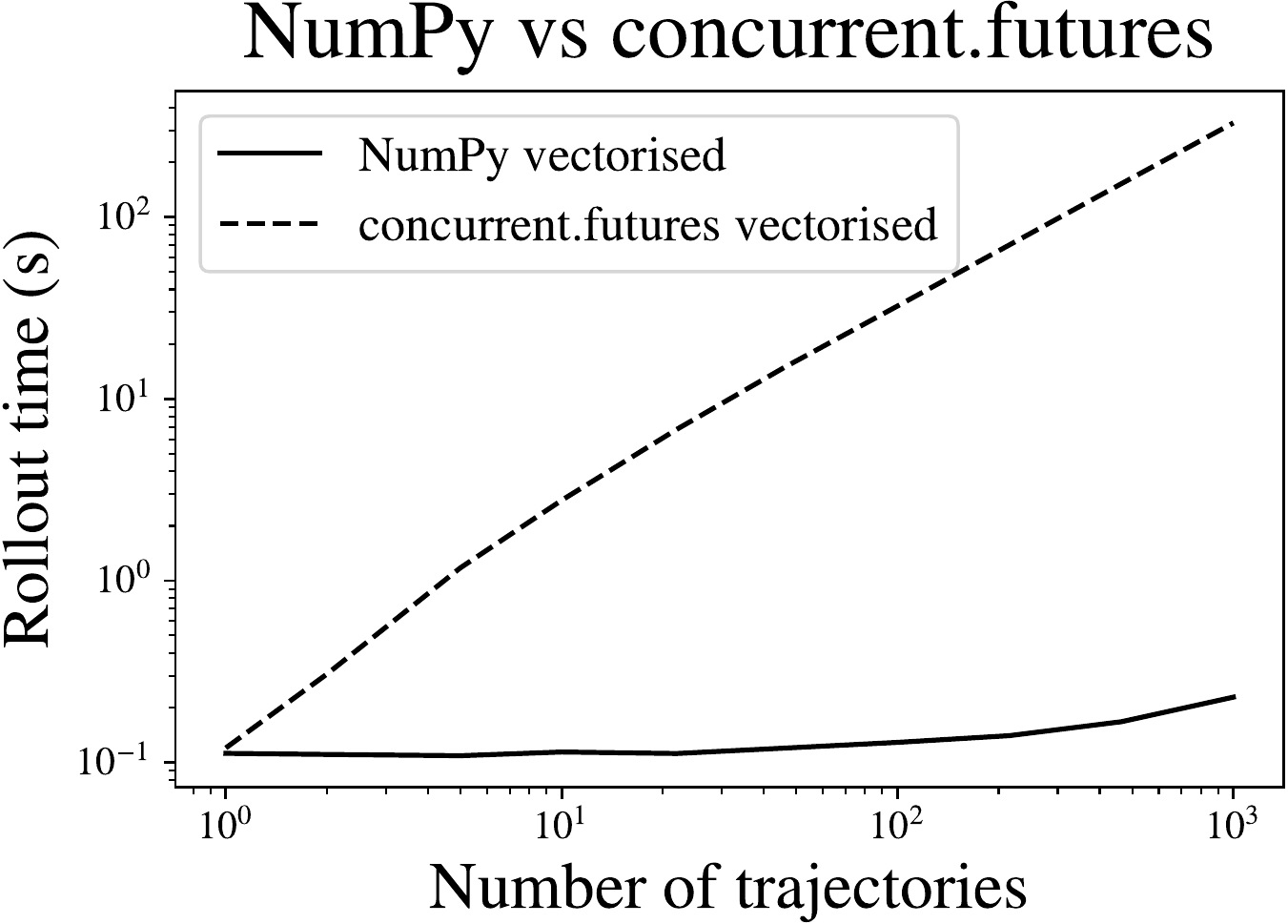}
\caption{Speedup of vectorization using NumPy.}
\label{fig:np-vs-concurrent}
\end{figure}

This approach enables deep reinforcement learning algorithms to scale well with computing resources. However, the structure of model-based market-making problems permits a much more efficient mode of parallelization -- namely that many trajectories can be simulated simultaneously using vector transformations from linear algebra. In particular, \mbtgym~uses NumPy arrays \cite{van2011numpy} as the data class for states, actions and rewards. A demonstration of the speedup that this approach offers over the multiprocessing approach is shown in Figure \ref{fig:np-vs-concurrent}. When rolling out 1000 trajectories, the multiprocessing approach takes 5 minutes and 30 seconds, whereas the NumPy approach takes 0.2 seconds.\footnote{Rollouts were generated using an AMD Ryzen 7 3800X 8-Core Processor with 16 threads and 64GB of RAM.}
 In practice, when training with a policy gradient approach we used 10,000 (or even 1,000,000) rollouts. 
 This would be prohibitively slow using multiprocessing.

\renewcommand{\arraystretch}{1.25}
\begin{table*}
\resizebox{\textwidth}{!}{
\begin{tabular}{|c|c|c|c|c|}
\hline
Name & Arrival process & Mid-price process & Actions & Reward function \\
\hline
Avellaneda and Stoikov \cite{avellaneda2008high} & 
\texttt{PoissonArrivalModel}	& 
\texttt{BrownianMotionMidpriceModel} &
limit   							 &
\texttt{ExponentialUtility} \\
Market-making with limit orders \cite[Section 10.2]{cartea2015algorithmic} &
\texttt{PoissonArrivalModel}	& 
\texttt{BrownianMotionMidpriceModel} &
limit & 
\texttt{RunningInventoryPenalty}\\
Market-making at the touch \cite[Section 10.2.2]{cartea2015algorithmic} &
\texttt{PoissonArrivalModel}	& 
\texttt{BrownianMotionMidpriceModel} &
touch & 
\texttt{RunningInventoryPenalty}\\
Cartea, Jaimungal, and Ricci (CJR)  \cite{CJR14} &
\texttt{HawkesArrivalProcess}	& 
\texttt{OuJumpDriftMidpriceModel} &
limit & 
\texttt{RunningInventoryPenalty}\\
 Gu\'eant, Lehalle, and Fernandez-Tapia\cite{gueant2013dealing} -- section 5.2 &
\texttt{PoissonArrivalModel}	& 
\texttt{BrownianMotionJumpMidpriceModel} &
limit   							 &
\texttt{ExponentialUtility} 
\\
Optimal execution with limit and market orders \cite[Section 8.4]{cartea2015algorithmic} &
\texttt{PoissonArrivalModel} &
\texttt{BrownianMotionMidpriceModel} &
limit\_and\_market & 
\texttt{RunningInventoryPenalty}\\
\hline
\end{tabular}
}
\caption{Examples of standard models from the literature implemented in \mbtgym}
\label{tab:standard}
\end{table*}
 
\vspace*{-0.25cm}

\begin{table*}
\resizebox{\textwidth}{!}{
\begin{tabular}{|c|c|c|c|c|}
\hline
Name & Arrival process & Mid-price process & Actions & Reward function \\
\hline
Avellaneda and Stoikov with Hawkes’ order flows &
\texttt{HawkesArrivalProcess}	& 
\texttt{BrownianMotionMidpriceModel} &
limit   							 &
\texttt{ExponentialUtility} \\
CJR-14 with pure jump price process &
\texttt{HawkesArrivalProcess}	& 
\texttt{OuJumpDriftMidpriceModel} &
touch & 
\texttt{RunningInventoryPenalty}\\
CJR-14 with exponential utility &
\texttt{HawkesArrivalProcess}	& 
\texttt{OuJumpDriftMidpriceModel} &
limit & 
\texttt{ExponentialUtility}\\
\hline
\end{tabular}
}
\caption{Examples of supported hybrid models}
\label{tab:hybrid}
\end{table*}

\section{Baseline agents}\label{sec:baseline}

We have implemented a number of baseline agents:

\begin{itemize}[leftmargin=0.3cm]
\item \texttt{RandomAgent} takes a random action in each step.
\item \texttt{FixedActionAgent} takes a fixed action in each step.
\item \texttt{FixedSpreadAgent} is a market-making agent that maintains a fixed spread (symmetric around the market mid-price).
\item \texttt{HumanAgent} allows a human to interact with the environment. 
\item \texttt{AvellanedaStoikovAgent} is optimal for~\cite{avellaneda2008high}.
\item \texttt{CarteaJaimungalAgent} is optimal for~\cite[Section 10.2]{cartea2015algorithmic}.

\end{itemize}

\noindent
These agents are useful for learning about models and environments. 
The \texttt{AvellanedaStoikovAgent}
and \texttt{CarteaJaimungalAgent} are useful for benchmarking solutions found by RL;
in Section~\ref{sec:training}, we give an example of training an agent 
for the \texttt{CarteaJaimungalAgent} setting, and then compare
it with the optimal baseline agent.


\section{A simple learning example}
\label{sec:training}
\begin{figure}[h]
\centering
\includegraphics[width=0.8\linewidth]{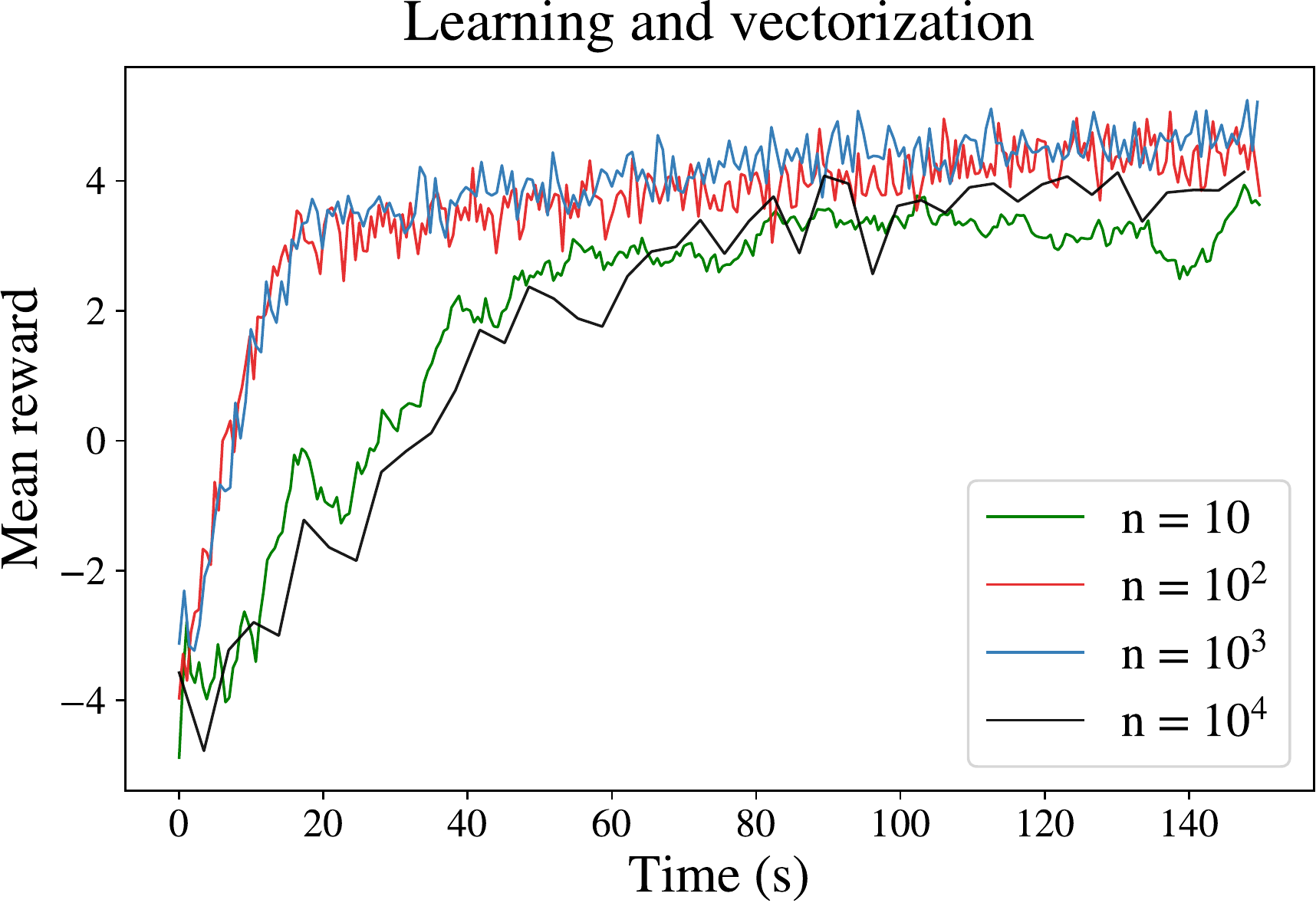}
\caption{The effect of different degrees of vectorization.}
\label{fig:learning-curves}
\end{figure}

\begin{figure}[h]
\centering
\includegraphics[width=0.8\linewidth]{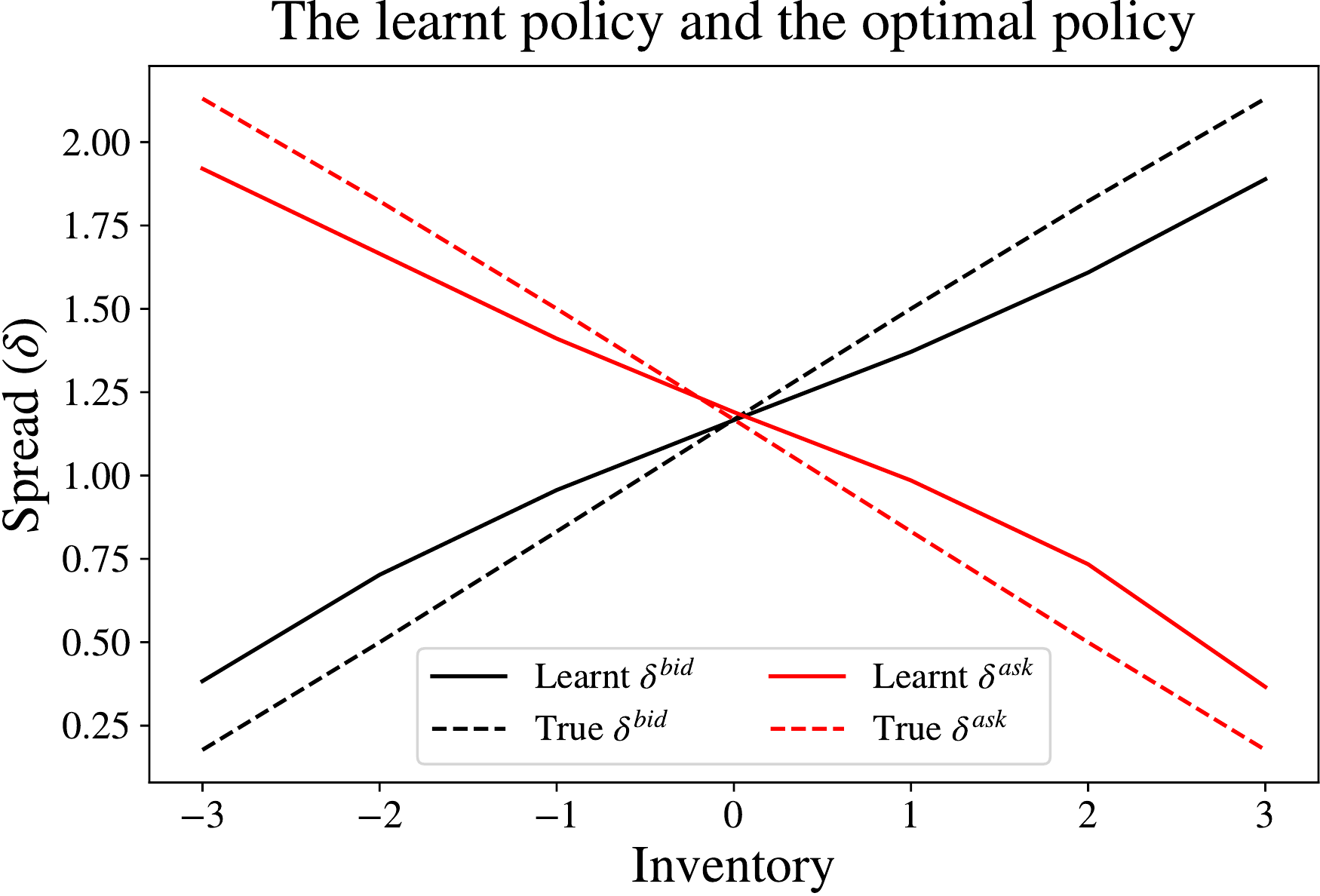}
\caption{Learnt PPO agent versus \texttt{CarteaJaimungalAgent}.}
\label{fig:learned-policy}
\end{figure}

As an example of training an agent using \mbtgym, we used proximal policy optimization\footnote{We used the StableBaselines3 implementation of PPO.} (PPO) \cite{SchulmanWDRK17} to train an agent to solve the market making problem considered in \cite[Section 10.2]{cartea2015algorithmic}. This problem is a suitable test bed for showing that this approach works since an explicit optimal solution is known. 
Figure \ref{fig:learning-curves} shows the evolution of mean rewards per episode against time for different degrees of vectorization and Figure \ref{fig:learned-policy} compares the learnt policy with the optimal policy from \cite[Section 10.2]{cartea2015algorithmic}.

In Figure \ref{fig:learning-curves} it is evident that if the degree of parallelization is too small (n=10 trajectories), then learning is unstable. Similarly, if number of trajectories rolled out in parallel is too large (n=10,000 trajectories), then the (wall-clock) time until convergence starts to increase. There is little difference in the training performance for n=100 and n=1000, but n=1000 outperforms n=100 slightly, with lower variance and convergence to higher mean rewards.

In Figure \ref{fig:learned-policy}, we see that the agent manages to learn the policy fairly well. The agent learns to skew their bid depth down and their ask depth up when they hold a negative inventory and vice-versa. This induces a mean reversion of the inventory around zero.

In addition to training a PPO agent, we provide an example in the code-base of solving the same problem using \textit{vanilla policy gradient}, a simpler policy gradient algorithm not in SB3, but implemented by us. To aid learning, the initial inventory can take random integer values in an interval to increase the states that the agent sees.

\section{Roadmap for further development}

\martin{I updated the new reward function names in all tables. I realised that table 3 is never referred to in the text. Do we really need it? (for the workshop paper). If so, we should briefly refer to it. Also I would replace the NAs for Mariariu-Patrichi and Pakkannen by touch and RunningInventoryPenalty.}

There are a number of extensions that the gym environment can accommodate in the
near future. The examples we give next are categorized into: (i) introducing
novel modelling assumptions about the dynamics of the stochastic processes
involved, (ii) incorporating the body of literature that deals with optimal
execution through trading speeds into the framework, and (iii) presenting
\mbtgym~working in conjunction with other RL libraries.

Regarding (i), our gym environment is readily suitable for a number of
extensions. We can explore the changes in optimal trading strategies once we
allow the fill probability model to be influenced by order arrivals, actions,
and fills themselves. For example, in an exponential fill probability model,
$\mathbb{P}[\text{Fill}] = e^{-\kappa\,\delta}$,
where $\delta\geq 0$ is the depth at which the market maker is posted and
$\kappa>0$ is the fill exponent, one could take $\kappa$ to be a stochastic
process that is affected by arrivals, actions, and fills.

One can also implement the action type \texttt{touch-and-market}, where the agent has control over 
whether to post at the best quotes, and also when to send a market order for a unit of the asset.


Other extensions can be more challenging to implement, for example, (a) granular
arrival of orders given by multi-dimensional Hawkes processes -- see Section 5
in \cite{abergel2020algorithmic}, or (b) allowing for latency in the framework
-- see \cite{gao2020optimal,cartea2021optimal}. Let us discuss (b) in a little more detail. In \cite{cartea2021optimal},
the time (in the agent's clock) between an order being sent to the exchange and its execution is exponentially distributed (random latency) or fixed (deterministic latency). Either of these two options can be implemented in our environment  by defining the update of the mid-price process in conjunction with a \texttt{LatencyProcess} class (inheriting from \texttt{StochasticProcess}) that defines the (possibly random) delay effects. Then, we update the mid-price process from $t$ to $t+\ell$ where $\ell$ is the latency, perform the transaction at $t+\ell$,  and then update the mid-price from $t+\ell$ to $t+\Delta$ where $\Delta$ is the time step -- note that we would assume that $\mathbb{P}[\ell\in(0,\Delta]] = 1$.
By doing so, we track latency effects internally, as opposed to exogenously defining it as an artificial cost.

In terms of (ii), we envisage the implementation of the action type \texttt{trading\_speed}, where the control is the trading speed at which the trader is buying or selling. Similar to the fill
exponent class, one would need to introduce the price impact class, that will
open the door to the study of optimal trading strategies under various price
impact functions or processes.
Further developments in this direction include the hybrid action types
\texttt{speed-and-limit}, or \texttt{speed-and-touch}, where the
agent trades at a chosen speed and simultaneously offers liquidity with the hope
of getting filled at better prices than those aggressively taken -- see
\cite{cartea2015optimal}.

For (iii) we would like to integrate
RLLib and RLax\footnote{\url{https://docs.ray.io/en/latest/rllib/index.html}; \url{https://github.com/deepmind/rlax}},
as alternatives to Stable Baselines.
Finally, a natural extension would be from single-agent RL to multi-agent
RL, e.g., to allow the training of RL agents that are robust across model parameters, e.g., as in~\cite{spooner2020robust}.


\section{Conclusion}

We presented~\mbtgym, a library of environments for applying RL to model-based LOB trading problems, 
along with a development roadmap.
We welcome contributions from the wider community. 




\bibliographystyle{ACM-Reference-Format}
\bibliography{papers}

\end{document}